\documentclass[12pt]{article}
\usepackage{epsfig}
\usepackage[T1]{fontenc}
\usepackage{graphics}

\usepackage[latin1]{inputenc}

\usepackage[english,french]{babel}

\usepackage{graphicx}

\usepackage{dcolumn}

\usepackage{amsmath,amssymb,amsfonts}

\begin{document}
\numberwithin{equation}{section}

\begin{flushleft}
\title*{\textbf{Superintegrable Systems with a Third Order Integrals of
Motion}}
\newline
\newline
Ian Marquette
\newline
D\'epartement de physique et Centre de recherche math\'ematiques,
Universit\'e de Montr\'eal,
\newline
C.P.6128, Succursale Centre-Ville, Montr\'eal, Qu\'ebec H3C 3J7,
Canada
\newline
ian.marquette@umontreal.ca
\newline
514-343-6111 (4072)
\newline
\newline
Pavel Winternitz
\newline
D\'epartement de math\'ematiques et de statistique et Centre de
recherche math\'ematiques,Universit\'e de Montr\'eal,
\newline
C.P.6128, Succursale Centre-Ville, Montr\'eal, Qu\'ebec H3C 3J7,
Canada
\newline
wintern@CRM.UMontreal.CA
\newline
514-343-7271
\newline
\newline
Two-dimensional superintegrable systems with one third order and
one lower order integral of motion are reviewed. The fact that
Hamiltonian systems with higher order integrals of motion are not
the same in classical and quantum mechanics is stressed. New
results on the use of classical and quantum third order integrals
are presented in Section 5 and 6.
\section{Introduction}
The purpose of this article is to present some new results and insights on superintegrable systems in classical and quantum mechanics, involving integrals of motion that are cubic in the momenta. We also present a review of this field with emphasis on the differences between "quadratic" and "cubic" integrability and superintegrability.
\newline
We recall that in classical mechanics a Hamiltonian system with Hamiltonian H and integrals of motion $X_{a}$
\newline
\begin{equation}
H=\frac{1}{2}g_{ik}p_{i}p_{k}+V(\vec{x},\vec{p}),\quad X_{a}=f_{a}(\vec{x},\vec{p}),\quad a=1,..., n-1
\end{equation}
\newline
is called completely integrable (or Liouville integrable) if it
allows n integrals of motion (including the Hamiltonian) that are
well defined functions on phase space, are in involution
$\{H,X_{a}\}_{p}=0$, $\{X_{a},X_{b}\}_{p}=0$, a,b=1,...,n-1 and
are functionally independent ($\{,\}_{p}$ is a Poisson bracket). A
system is superintegrable if it is integrable and allows further
integrals of motion $Y_{b}(\vec{x},\vec{p})$, $\{H,Y_{b}\}_{p}=0$,
b=n,n+1,...,n+k that are also well defined functions on phase
space and the integrals$\{H,X_{1},...,X_{n-1},Y_{n},...,Y_{n+k}\}$
are functionally independent. A system is maximally
superintegrable if the set contains 2n-1 functions, minimally
superintegrable if it contains n+1 such integrals. The integrals
$Y_{b}$ are not required to be in evolution with
$X_{1}$,...$X_{n-1}$, nor with each other.
\newline
The same definitions apply in quantum mechanics but
$\{H,X_{a},Y_{b}\}$ are well defined quantum mechanical operators,
assumed to form an algebraically independent set.
\newline
The best known examples of (maximally) superintegrable systems are
the Kepler-Coulomb [1,2] system $V(\vec{x})=\frac{\alpha}{r}$ and
the harmonic oscillator $V(\vec{x})=\alpha r^{2}$ [3,4]. It
follows from Bertrand's theorem [5,6] that these are the only two
spherically symmetric examples in Euclidean space.
\newline
In both cases the integrals of motion are first or second order
polynomials in the momenta. The Hamiltonians and integrals are the
same in classical and quantum mechanics (after a possible
symmetrization).
\newline
A systematic search for superintegrable systems in two-dimensional
Euclidean space $E_{2}$ was started some time ago [7,8].
Rotational symmetry was not imposed. The integrals were assumed to
be second order polynomials in the momenta. Thus, the Ansatz (in
quantum mechanics) was
\newline
\begin{equation}
H=\frac{1}{2}(\vec{p})^{2}+V(\vec{x}) \quad ,    \quad  \quad
\vec{x}=(x_{1},x_{2})
\end{equation}
\[X_{a}=\sum_{i,k=1}^{2}\{f_{a}^{ik}(\vec{x}),p_{i}p_{k}\}+\sum_{i=1}^{2}g_{a}^{i}(\vec{x})p_{i}+\phi_{a}(\vec{x}) \]
where $\{,\}$ is an anticommutator and we put
\newline
\begin{equation}
p_{j}=-i\hbar \frac{\partial}{\partial x_{j}},\quad L_{3}=x_{2}p_{1}-x_{1}p_{2}
\end{equation}
\newline
The commutativity requirement $[H,X_{a}]$=0 implies that $X_{a}$
must have the following form:
\newline
\begin{equation}
X=aL_{3}^{2}+b(L_{3}p_{1}+p_{1}L_{3})+c(L_{3}p_{2}+p_{2}L_{3})+d(p_{1}^{2}-p_{2}^{2})+2fp_{1}p_{2}+\phi(x,y),
\end{equation}
\newline
where a,...,f are constants (or X can be first order operator).
\newline
Using transformations from the Euclidean group E(2) (they leave
the form of the Hamiltonian (1.2) invariant) we can transform the
integral (1.4) into one of four standard forms. If one such
operator X exists then the potential $V(x_{1},x_{2})$ will allow
the separation of variables in cartesian, polar, parabolic or
elliptic coordinates, respectively. If two operators
$\{X_{1},X_{2}\}$, commuting with H exist, we obtain four families
of superintegrable potentials, [7,8], namely
\newline
\begin{equation}
V_{I}=\alpha(x^{2}+y^{2})+\frac{\beta}{x^{2}}+\frac{\gamma}{y^{2}},\quad V_{II}=\alpha(x^{2}+4y^{2})+\frac{\beta}{x^{2}}+\gamma y
\end{equation}
\[V_{III}=\frac{\alpha}{r}+\frac{1}{r^{2}}(\frac{\alpha}{1+cos(\phi)}+\frac{\beta}{1-cos(\phi)}),\quad V_{IV}=\frac{\alpha}{r}+\frac{1}{r}(\beta cos(\frac{\phi}{2})+\gamma sin(\frac{\phi}{2}))               \]
Each of these potentials is multiseparable, i.e. allows the
separation of variables in the Schrödinger equation (and in the
Hamilton-Jacobi equation) in at least two systems of coordinates.
\newline
A similar search for superintegrable systems in 3 dimensional
Euclidean space $E_{3}$ gave a complete classification of all
"quadratically superintegrable" systems in $E_{3}$ [9,10].
\newline
Since then many results have been obtained on superintegrable
systems in $E_{n}$ for n arbitrary and for two dimensional spaces
of constant curvature, so called Darboux spaces (2 dimensional
spaces with non constant curvature allowing at least two killing
tensors), for complex spaces, etc [11,...,15].
\newline
The overall picture that emerges is that superintegrable systems
with quadratic (or linear) integrals of motion have the following
properties that make them interesting from the point of view of
physics and mathematics:
\newline
\newline
1. All finite classical trajectories are closed (more generally,
they are constrained to an n-k dimensional manifold in phase space
[16]).
\newline
2. They are all multiseparable (in classical and quantum
mechanics).
\newline
3. Their quantum energy levels are degenerate.
\newline
4. The Schrödinger equation for quadratically superintegrable
systems in $E_{2}$ is exactly solvable and it has been conjectured
that this is true in general [17]. The conjecture is true in all
cases so far studied.
\newline
5. The same potentials are superintegrable in classical and
quantum mechanics.
\newline
6. The integrals of motion form non-Abelian algebras under
commutation, or Poisson commutation, respectively. These algebras
may be finite dimensional, or infinite dimensional Lie algebras.
The infinite-dimensional ones have a special structure and it is
often advantagous to view them as polynomial algebras, more
specifically quadratic ones [18,19,20].
\newline
It would be interesting to establish which of the above properties
hold for all superintegrable systems and which are restricted to
Hamiltonians of the form (1.1) with quadratic integrals of motion.
Other systems that have been studied include velocity dependent
potentials [21-25] and particles with spin [26].In this article we
review the case of a scalar potential as in (1.1) but with third
order integrals of motion. We also present some new results.
\section{Integrability with third order integral of motion.}
In 1935 J. Drach published two articles on two-dimensional
Hamiltonian systems with third order integrals of motion [27]. The
main feature of these papers are
\newline
1. The author found 10 such integrable potentials, each depending
on 3 constants, but not involving any arbitrary functions.
\newline
2. The results were obtained in classical mechanics and are in
general not valid in quantum mechanics.
\newline
3. The space considered was a complex Euclidean space
$E_{2}(\mathbb{C})$.
\newline
4. The articles are very short and "condensed". Basically they are
research announcements, but no detailed follow up articles were
published.
\newline
5. Some assumptions were made in the calculation so it is not
clear whether the obtained list is complete.
\newline
Much more recently it was shown by Ra\~nada [28] and Tsiganov [29]
that 7 of the 10 Drach potentials are actually "reducible". The 7
corresponding Hamiltonian systems are superintegrable, in that
they allow two second order integrals. Their Poisson commutator is
the third order integral found by Drach.
\newline
Since then many examples of systems with third and higher order
integrals have been published, mainly in classical
mechanics[30,...,34].
\newline
In 1998 J.Hietarinta published a remarkable article "Pure quantum
integrability" [35] in which he showed that potentials exist that
are integrable in quantum mechanics but have free motion as their
classical limit. The  integrals of motion are third or higher
order polynomials in the momenta and the potentials have the form
$V(x,y)=\hbar^{2}f(x,y)$ where $\hbar$ is the Planck constant and
f(x,y) remains finite in the limit $\hbar \rightarrow 0$.
\newline
A systematic search for integrable and superintegrable systems in
both classical and quantum mechanics was initiated in Ref 36. The
problem was formulated in quantum mechanics, the classical case
was considered as a limit for $\hbar \rightarrow 0$. The
Hamiltonian and first integral have the form
\newline
\begin{equation}
H=\frac{1}{2}(p_{1}^{2}+p_{2}^{2})+V(x_{1},x_{2})
\end{equation}
\newline
\begin{equation}
X=\frac{1}{2}\sum_{j,k}f_{jk}(x_{1},x_{2})p_{1}^{j}p_{2}^{k} \quad
0 \leq j+k \leq 3
\end{equation}
\newline
where $p_{j}$ are as in (1.3) and the potential V and functions
$f_{jk}$ are to be determined. The commutativity requirement
[H,X]=0 implies that for any potential the third order integral X
will have the form
\newline
\begin{equation}
X=\sum_{i+j+k=3}A_{ijk}\{L_{3}^{i},p_{1}^{j}p_{2}^{k}\}+\{g_{1}(x_{1},x_{2}),p_{1}\}+\{g_{2}(x_{1},x_{2}),p_{2}\}
\end{equation}
\newline
where $\{,\}$ is an anticommutator, $L_{3}$ is the angular
momentum, as in eq. (1.3). The constants $A_{ijk}$ and functions
V, $g_{1}$ and $g_{2}$ are subject to further determining
equations:
\begin{equation}
(g_{1})_{x_{1}}=3f_{1}(x_{2})V_{x_{1}}+f_{2}(x_{1},x_{2})V_{x_{2}} ,\quad (g_{2})_{x_{2}}=f_{3}(x_{1},x_{2})V_{x_{1}}+3f_{4}(x_{1})V_{x_{2}}
\end{equation}
\[   (g_{1})_{x_{2}}+(g_{2})_{x_{1}}=2(f_{2}(x_{1},x_{2})V_{x_{1}}+f_{3}(x_{1},x_{2})V_{x_{2}}                    \]
\newline
and
\newline
\begin{equation}
g_{1}V_{x_{1}}+g_{2}V_{x_{2}}=\frac{\hbar^{2}}{4}(f_{1}V_{x_{1}x_{1}x_{1}}+f_{2}V_{x_{1}x_{1}x_{2}}+f_{3}V_{x_{1}x_{2}x_{2}}+f_{4}V_{x_{2}x_{2}x_{2}}  )
\end{equation}
\[+8A_{300}(x_{1}V_{x_{2}}-x_{2}V_{x_{1}})+2(A_{210}V_{x_{1}}+A_{201}V_{x_{2}})\]
The functions $f_{i}$ are defined in terms of the constants $A_{ijk}$ as
\newline
\begin{equation}
f_{1}=-A_{300}x_{2}^{3}+A_{210}x_{2}^{2}-A_{120}x_{2}+A_{030}
\end{equation}
\[ f_{2}=3A_{300}x_{1}x_{2}^{2}-2A_{210}x_{1}x_{2}+A_{201}x_{2}^{2}+A_{120}x_{1}-A_{111}x_{2}+A_{021}                \]
\[ f_{3}=-3A_{300}x_{1}^{2}x_{2}+A_{210}x_{1}^{2}-2A_{201}x_{1}x_{2}+A_{111}x_{1}-A_{102}x_{2}+A_{012}                \]
\[f_{4}=A_{300}x_{1}^{3}+A_{210}x_{1}^{2}+A_{102}x_{1}+A_{003}                  \]
\newline
We can sum up the results on third order integrability as follows.
\newline
1. The leading part of the integral X lies in the enveloping
algebra of the Euclidean Lie algebra e(2).
\newline
2. Even and odd terms in X commute with H separately. We required
the existence of a third order integral. A second order one of the
form (1.4) may, or may not exist.
\newline
3. The 10 constants and 3 functions $g_{1},g_{2}$ and V are to be
determined from the (overdetermined) systems of 4 equations (2.4)
and (2.5). The fact that the system is overdetermined implies the
V(x,y) will satisfy some compatibility conditions. Indeed, one can
obtain 1 linear third order equation for V and 3 nonlinear ones
[36]. The existence of a third order integral is hence much more
constraining than the existence of a second order one ( in
agreement with Drach's results).
\newline
4. The classical and quantum cases differ! Indeed, the determining
equation (2.4) are the same in both cases, but eq. (2.5) involves
the Planck constant $\hbar$. In the classical limit $\hbar
\rightarrow 0$ it simplifies greatly ( the right hand side goes to
zero).
\newline
5. Instead of attempting to find a general solution we turn to a
simpler problem, namely that of superintegrability. We shall
consider potentials V(x,y) for which two integrals of motion
exist, one third order one and one that is either a first, or
second order polynomial in the momenta.
\section{Systems with one first order and one third order integral}
Let us now assume that the Hamiltonian (2.1) allows a first order
integral of motion Y. This means that the potential has a
geometric symmetry: it is invariant either under rotations, or
translations in one direction. With no loss of generality we can
consider just two representative cases:
\newline
\begin{eqnarray}
&& 1)V=V(r), \quad Y=L_{3},\qquad r=\sqrt{x^2+y^2,}\\
&& 2) V=V(x), \quad Y=P_{2}. \nonumber
\end{eqnarray}
\newline
 Eq. (2.4) and (2.5) for the existence of a
third order integral X greatly simplify in both cases.
\newline
Let us first consider classical mechanics, i.e. $\hbar=0$ in eq.
(2.5). Solving the systems (2.4) and (2.5) for the above
potentials, we obtain
\newline
\begin{equation}
V_{1}=\frac{\alpha}{r}, \quad V_{2}=\alpha r^{2},\quad  V_{3}=ax ,
\quad V_{4}=\frac{a}{x^{2}}.
\end{equation}
\newline
All of these potential are well know [1,...,8,37] as quadratically
superintegrable. Indeed, the third order integral in all these
cases is the Poisson commutator of a second order integral with
the square of the first order one.
\newline
In quantum mechanics we again obtain the potentials (3.2) but in
addition we obtain a potential V(x) satisfying
\newline
\begin{equation}
\hbar^{2}(V''(x))^{2}=4V^{3}+\alpha V^{2}+\beta V + \gamma \equiv 4(V-A_{1})(V-A_{2})(V-A_{3})
\end{equation}
\newline
where $\alpha$,$\beta$ and $\gamma$ (and $A_{1}$,$A_{2}$,$A_{3}$)
are constants. Eq. (3.3) is solved in terms of elliptic functions.
Depending on properties of the roots $A_{i}$ and on the initial
conditions, we obtain 3 types of solutions:
\begin{equation}
V=(\hbar \omega)^{2}k^{2}sn^{2}(\omega x,k),\quad V=\frac{(\hbar \omega)^{2}}{2(cn(\omega x ,k)+1)}
\end{equation}
\[V=(\hbar \omega)^{2}\frac{1}{sn^{2}(\omega x,k)}                \]
where $sn(\omega x,k)$ and $cn(\omega x,k)$ are Jacobi elliptic
functions. If two of the roots $A_{i}$ coincide, we get potentials
expressed in terms of elementary functions
\newline
\begin{equation}
V=\frac{(\hbar \omega)^{2}}{(cosh(\omega x))^{2}} ,\quad V=\frac{(\hbar \omega)^{2}}{(sinh(\omega x))^{2}}
\end{equation}
\[V=\frac{(\hbar \omega)^{2}}{sin^{2}(\omega x)}         \]
\newline
We see explicitly that for $\hbar \rightarrow 0$ all these
potentials vanish. The potentials (3.4) and (3.5) are
"irreducible" and genuinely superintegrable. The two algebraically
independent integrals are
\newline
\begin{equation}
Y=P_{2},\quad
X=\{L_{3},p_{1}^{2}\}+\{(\sigma-3V)y,p_{1}\}+\{-\sigma x+2xV+\int
V(x)dx,p_{2}\}  ,
\end{equation}
$\sigma=A_{1}+A_{2}+A_{3}$
\newline
Even though the potentials V(x) in (3.4) and (3.5) are
one-dimensional (depend on x alone), their superintegrability is a
two-dimensional phenomenon, since the integral X involves
rotations $L_{3}$.
\section{Systems with one second and one third order integral}
Let us now assume that the Hamiltonian (2.1) allows one second
order integral of motion Y. This integral will have the form
(1.4). As mentioned in the Introduction, 4 classes of such
Hamiltonians exist [7,8]. Here we shall restrict to one of the 4
classes, namely to potentials allowing the separation of variables
in cartesian coordinates. Thus we have
\newline
\begin{equation}
Y=\frac{1}{2}(p_{1}^{2}-p_{2}^{2})+V_{1}(x)-V_{2}(y),\quad
V(x,y)=V_{1}(x)+V_{2}(y).
\end{equation}
\newline
We substitute the above potential into the determining equations
(2.4) and (2.5) in order to determine potentials that also allow a
third order integral. The obtained system of equations is quite
manageable and was analyzed by S.Gravel [38].
\newline
The results in the classical and quantum cases are very different.
To illustrate the difference, let first consider an example. One
particular solution of the determining system (2.4), (2.5) is the
potential
\newline
\begin{equation}
V(x,y)=\frac{\omega^{2}}{2}y^{2}+V(x)
\end{equation}
\newline
where V(x) satisfies a fourth order nonlinear ordinary differential equation
\newline
\begin{equation}
\hbar^{2}V^{''''}=12\omega^{2}xV'+6(V^{2})''-2\omega^{2}x^{2}V''+2\omega^{2}x^{2}.
\end{equation}
\newline
We mention that this equation is also obtained as a nonclassical
reduction of the Bousinesq equation [39,40].For $\hbar \neq 0$
this equation is solved in terms of the Painlev\'e transcendent [41]
$P_{IV}(x,-\frac{4\omega^{2}}{\hbar})$. The classical limit is
singular ($\hbar \rightarrow 0$ reduces the order of the equation
). In the classical case we can reduce to quadratures and obtain a
quartic algebraic equation for V(x):
\newline
\begin{equation}
-9V^{4}(x)+14\omega^{2}x^{2}V^{3}(x)+(6d-\frac{15}{2}\omega^{4}x^{4})V^{2}(x)+
\end{equation}
\[(\frac{3}{2}\omega^{6}x^{6}-2d\omega^{2}x^{2})V(x)+cx^{2}-d^{2}-d\frac{\omega^{2}}{2}x^{4}-\frac{1}{16}\omega^{8}x^{8}=0              \]
For special values of the constants a,...,d,$\omega$ namely
\newline
\begin{equation}
c=\frac{8\omega^{8}b^{3}}{36},\quad d=\frac{\omega^{4}b^{2}}{3^{3}}
\end{equation}
\newline
eq. (4.4) has a double root and we obtain
\newline
\begin{equation}
V_{1,2}=\frac{\omega^{2}}{18}(2b+5x^{2}\pm 4x\sqrt{b+x^{2}} ),
\quad
V_{3}=V_{4}=\frac{\omega^{2}}{2}x^{2}-\frac{\omega^{2}b}{3^{3}}
\end{equation}
A complete analysis of eq. (2.4) and (2.5) is given in Ref. 38.
All together 8 classical superintegrable systems that are
separable in cartesian coordinates and allow at least one third
order integral of motion exist. Three of them are reducible (the
harmonic oscillator and potentials $V_{I}$ and $V_{II}$ of eq.
(1.5))
\newline
\newline
The 5 irreducible classical potentials are
\newline
$V=\frac{\omega^{2}}{2}(9x^{2} + y^{2})$
\newline
$V=\beta_{1}^{2}\sqrt{|x|} + \beta_{2}^{2}\sqrt{|y|}$
\newline
$V=a^{2}|y| + b^{2}\sqrt{|x|}$
\newline
$V=\frac{\omega^{2}}{2}y^{2} + V(x)$
\newline
$V=a|y| + f(x)$,
\newline
where V(x) satisfies eq. (4.4) and f(x) satisfies $f^{3} -
2bxf^{2} + b^{2}x^{4}f-d=0$.
\newline
\newline
The quantum case is much richer: 21 superintegrable cases of the
considered type exist, 13 of them irreducible. The potentials are
expressed in terms of rational functions in 6 cases, elliptic
functions in 2 cases and Painlev\'e transcendents [41] $P_{I}$,$P_{II}$
and $P_{IV}$ in 5 cases. Let us just present the irreducible
potentials.
\newline
\newline
Rational function potentials:
\newline
$V=\hbar^{2}[
\frac{x^{2}+y^{2}}{8a^{4}} +
\frac{1}{(x-a)^{2}}+\frac{1}{(x+a)^{2}}]$
\newline
$V=\hbar^{2}[\frac{1}{8a^{4}}(x^{2}+y^{2})+\frac{1}{y^{2}}+\frac{1}{(x+a)^{2}}+\frac{1}{(x-a)^{2}}
]  $
\newline
$V=\hbar^{2}[\frac{1}{8a^{4}}(x^{2}+y^{2})+\frac{1}{(y+a)^{2}}+\frac{1}{(y-a)^{2}}
+\frac{1}{(x+a)^{2}}+\frac{1}{(x-a)^{2}} ]   $
\newline
$V=\frac{\omega^{2}}{2}(9x^{2} + y^{2})  $
\newline
$V=\frac{\omega^{2}}{2}(9x^{2} + y^{2})+\frac{\hbar^{2}}{y^{2}}    $
\newline
$V=\hbar^{2}[
\frac{9x^{2}+y^{2}}{8a^{4}} +
\frac{1}{(y-a)^{2}}+\frac{1}{(y+a)^{2}}]$
\newline
\newline
Elliptic function potentials:
\newline
$V=\hbar^{2}P(y)+V(x)$, V(x) is arbitrary
\newline
$V=\hbar^{2}(P(y)+P(x))$, P(x) is a Weierstrass elliptic function
\newline
\newline
Painlev\'e transcendent potentials:
\newline
$V=\hbar^{2}(\omega_{1}^{2}P_{I}(\omega_{1}x)+\omega_{2}^{2}P_{I}(\omega_{2}y))$
\newline
$V=ay+\hbar^{2}\omega_{1}^{2}P_{I}(\omega_{1}x)$
\newline
$V=bx+ay+(2\hbar b)^{\frac{2}{3}}P_{II}^{2}((\frac{2b}{\hbar^{2}})^{\frac{1}{3}}x)$
\newline
$V=ay+(2\hbar^{2} b^{2})^{\frac{1}{3}}(P'_{II}((\frac{-4b}{\hbar^{2}})^{\frac{1}{3}}x)+P_{II}^{2}((\frac{-4b}{\hbar^{2}})^{\frac{1}{3}}x)                    )$
\newline
$V=a(x^{2}+y^{2})+\frac{\hbar^{2}}{2}(P'_{IV}(\frac{-8a}{\hbar^{2}}x)-\frac{1}{2}\sqrt{8a}P_{IV}^{2}(\frac{-8a}{\hbar^{2}}x)-\frac{1}{2}\sqrt{8a}xP_{IV}(\frac{-8a}{\hbar^{2}}x)               )                   $
\newline
\newline
The occurence of Painlev\'e transcendents as superintegrable
potentials seems somewhat surprising. It is less so once we
remember the relation between the Schrödinger equation and the
Korteweg-de Vries equation [42]. Solutions of the KdV include
Painlev\'e transcendents. Painlev\'e transcendent potentials have
already occured in other contexts [43,44,45].
\section{Trajectories for the classical systems}
The trajectories for all 8 superintegrable classical potentials
were presented in Ref. 46. The trajectories can be obtained by
integrating the equations of motion and then eliminating the time
variable from the result. Alternatively, the trajectories can be
obtained directly from the integrals of motion. Indeed, in all
cases we have
\newline
\begin{equation}
\frac{1}{2}p_{1}^{2}+f(x)=E_{1},\quad \frac{1}{2}p_{2}^{2}+g(x)=E_{2}
\end{equation}
\begin{equation}
X=\mu p_{1}^{3}+\nu p_{1}^{2}p_{2}+\rho p_{1}p_{2}^{2}+\sigma
p_{2}^{3}+\phi p_{1}+\psi p_{2}=K
\end{equation}
where $\mu$,...,$\sigma$ are known low order polynomials in x and
y, f(x),g(x),$\phi(x,y)$ and $\psi(x,y)$ are known functions and
$E_{1}$,$E_{2}$ and K are constants depending on the initial
conditions.
\newline
From eq. (5.1) we express $P_{1}$ and $P_{2}$ in terms of
$E_{1}$,$E_{2}$,f(x) and g(y) and substitute into eq (5.2). This
directly gives us an equation for the trajectory (not necessarily
in a convenient form). If the functions f(x) and g(y) satisfy f(x)
$\geq$ 0, g(y) $\geq$ 0, for $x^{2}$ > $x_{0}$, $y^{2}$ > $y_{0}$
for some constants $x_{0}$ and $y_{0}$, then the motion will be
bounded.
\newline
From the figures of Ref 46 we see that the finite trajectories are
all closed, as predicted by the general theory.
\section{Algebras of the quantum integrals of motion}
The integrals of motion in quantum mechanics form associative
polynomial (cubic) algebras with respect to Lie commutation. The
commutator relations in all cases are of the form
\newline
\begin{equation}
[A,B]=C,\quad [A,C]=\alpha B,\quad [B,C]=\beta A^{3}+\gamma A^{2}+\delta A +\epsilon
\end{equation}
\newline
where $\gamma$, $\delta$ and $\epsilon$ are polynomials in H (for
the classical cubic algebras of Poisson brackets see Ref. 46).
\newline
Inspired by the work of Daskaloyannis and al. on quadratic
algebras [18,47,48] we construct a realization of the cubic
algebra (6.1) using a deformed oscillator algebra and Fock space
basis $\{b^{t},b,N\}$. We put
\newline
\begin{equation}
[N,b^{t}]=b^{t} ,\quad [N,b]=-b ,\quad b^{t}b=\Phi(N) ,\quad
bb^{t}=\Phi(N+1)
\end{equation}
\newline
where $\Phi(N)$ is the so called "structure function". To
construct representations of the cubic algebra we impose
conditions on $\Phi(N)$. We require that a natural number p should
exist, such that
\begin{equation}
\Phi(p+1)=0
\end{equation}
and further we impose
\newline
\begin{equation}
\Phi(0)=0,\quad \Phi(x) > 0,\quad  x > 0
\end{equation}
\newline
We put
\begin{equation}
A=A(N),\quad B=b(N)+b^{t}\rho(N)+\rho(N)b
\end{equation}
\newline
where the functions A(N),b(N),$\rho(N)$ and $\Phi(N)$ are to be
determined from the cubic algebra (6.1).
\newline
The next step is to construct a finite dimensional representation of this algebra with A, N and K (the Casimir operator) diagonal. We obtain a realization of a parafermionic oscillator
\newline
\[N|k,n>=n|k,n>, \quad K|k,n>=k|k,n>\]
\[A|k,n>=A(k,n)|k,n>\]
\begin{equation}
A|k,n>=\sqrt{\delta}(n+u)|k,n>
\end{equation}
\[\Phi(0,u,k)=0,\quad \Phi(p+1,u,k)=0\]
A complete discussion of this approach to quantum superintegrable
systems with cubic integrals of motion will be presented
elsewhere.
\newline
Here we just present one example showing how one can obtain energy
spectra. Let us consider one of the rational quantum potentials of
Section 4, namely
\newline
\begin{equation}
V=\hbar^{2}[
\frac{x^{2}+y^{2}}{8a^{4}} +
\frac{1}{(x-a)^{2}}+\frac{1}{(x+a)^{2}}]
\end{equation}
\newline
The Casimir operator K and structure function $\Phi(x)$ are in this case found to be
\newline
\begin{equation}
 K = -16\hbar^{2}H^{4} + 32\frac{\hbar^{4}}{a^{2}}H^{3} +
16\frac{\hbar^{6}}{a^{4}}H^{2} - 40\frac{\hbar^{8}}{a^{6}}H -
3\frac{\hbar^{10}}{a^{8}}\quad .
\end{equation}
\newline
\begin{equation}
\Phi(x)=(\frac{-\hbar^{8}}{a^{4}})(x+u -
(\frac{-a^{2}E}{\hbar^{2}}-\frac{1}{2}))(x+u -
(\frac{a^{2}E}{\hbar^{2}}+\frac{1}{2}))(x+u -
(\frac{-a^{2}E}{\hbar^{2}}+\frac{3}{2}))(x+u -
(\frac{-a^{2}E}{\hbar^{2}}+\frac{5}{2}))
\end{equation}
\newline
The condition $\Phi(0,u,k)=0$ is now an equation for u and
$\Phi(p+1,u,k)=0$ provides the energy spectrum. In the case of the
potential (6.7) the constant a can be real, or pure imaginary
(V(x,y) is real in both cases). The energy spectrum we obtain is
\newline
\[E=\frac{\hbar(p+2)}{2a_{0}^{2}}  \quad \quad  a=ia_{0},\quad a_{0}\in \mathbb{R},\quad p\in \mathbb{N}\]
\[E=\frac{\hbar (p+3)}{2a^{2}}  \quad \quad  a \in \mathbb{R},\quad p\in \mathbb{N}\]
\section{Conclusion}
The main conclusion that we draw at this stage is that going
beyond quadratic integrability for scalar potentials opens new
horizons and poses new questions.
\newline
One of the questions is: What does one do with integrals of motion
that do not lead to the separation of variable in the equations of
motion ? A partial answer is given in this article and is very
different in classical and quantum mechanics. We have seen that
the integrals can be used to calculate trajectories directly
without solving the classical equation of motion. In quantum
mechanics one uses the algebra of integrals of motion to obtain
information about energy spectra and wave functions.
\newline
\newline
\textbf{Acknowledgments} The research of P.W. is partially
supported by a research grant from NSERC of Canada. This article
was written while he was on sabbatical leave at ENEA, Frascati,
Italy and Universita di Roma Tre. He thanks G.Dattoli and D.Levi
for hospitality. The research of I.M. is supported by a doctoral
research scholarship from FQRNT of Quebec. This article was
written while he was visiting the Universita di Roma Tre. He
thanks D.Levi for his hospitality.
\section{\textbf{References}}
1. V.Fock, Z.Phys. 98, 145-154 (1935).
\newline
2. V.Bargmann, Z.Phys. 99, 576-582 (1936).
\newline
3. J.M.Jauch and E.L.Hill, Phys.Rev. 57, 641-645 (1940).
\newline
4. M.Moshinsky and Yu.F.Smirnov, The Harmonic Oscillator In Modern
Physics, (Harwood, Amsterdam, 1966).
\newline
5. J.Bertrand, C.R. Acad. Sci. III, 77, 849-853 (1873).
\newline
6. H.Goldstein, Classical Mechanics, (Addison-Wesley, Reading, MA, 1990)
\newline
7. J.Fris, V.Mandrosov, Ya.A.Smorodinsky, M.Uhlir and P.Winternitz, Phys.Lett. 16, 354-356 (1965).
\newline
8. P.Winternitz, Ya.A.Smorodinsky, M.Uhlir and I.Fris, Yad.Fiz. 4,
625-635 (1966). (English translation in Sov. J.Nucl.Phys. 4,
444-450 (1967)).
\newline
9. A.Makarov, Kh. Valiev, Ya.A.Smorodinsky and P.Winternitz, Nuovo Cim. A52, 1061-1084 (1967).
\newline
10. N.W.Evans, Phys.Rev. A41, 5666-5676 (1990), J.Math.Phys. 32,
3369-3375 (1991).
\newline
11. E.G.Kalnins, J.M.Kress, W.Miller Jr and P.Winternitz,
J.Math.Phys. 44(12) 5811-5848 (2003).
\newline
12. E.G.Kalnins, W.Miller Jr and G.S.Pogosyan, J.Math.Phys. A34,
4705-4720 (2001).
\newline
13. E.G.Kalnins, J.M.Kress and W.Miller Jr, J.Math.Phys. 46,
053509 (2005), 46, 053510 (2005), 46, 103507 (2005), 47, 043514
(2006), 47, 043514 (2006).
\newline
14. E.G.Kalnins, W.Miller Jr and G.S.Pogosyan, J.Math.Phys. 47,
033502.1-30 (2006), 48, 023503.1-20 (2007).
\newline
15.M.A.Rodriguez and P.Winternitz, J.Math.Phys. 43(3), 1309-1322
(2002).
\newline
16. N.N.Nekhoroshev, Trans. Moscow Math.Soc 26, 180 (1972).
\newline
17. P.Tempesta, A.V.Turbiner and P.Winternitz, J.Math.Phys. 42,
4248 (2001).
\newline
18. C.Daskaloyannis, J.Math.Phys. 42, 110 (2001).
\newline
19. Ya.I.Granovskii, A.S.Zhedanov and I.M.Lutzenko, J.Phys. A24,
3887 (1991).
\newline
20. P.Letourneau and L.Vinet, Ann.Phys. 243, 144 (1995).
\newline
21. B.Dorizzi, B.Grammaticos, A.Ramani and P.Winternitz,
J.Math.Phys. 26, 3070-3079 (1985).
\newline
22. E.McSween and P.Winternitz, J.Math.Phys. 41, 2957-2967 (2000).
\newline
23 J.B\'erub\'e and P.Winternitz, J.Math.Phys. 45, 1959-1973
(2004).
\newline
24. F.Charest, C.Hudon and P.Winternitz, J.Math.Phys. 48,
012105.1-16 (2007).
\newline
25. G.Pucacco and K.Rosquist, J.Math.Phys. 46, 012701 (2005).
\newline
26. P.Winternitz and I.Yurdusen, J.Math.Phys. 47(10), 103509
(2006) and ArXiv math-ph.:0711.0753 (2007).
\newline
27. J.Drach, C.R.Acad.Sci.III, 200, 22 (1935), 200, 599 (1935).
\newline
28. M.F.Ra\~nada, J.Math.Phys. 38, 4165-4178 (1997).
\newline
29. A.V.Tsiganov, J.Math.Phys. A33, 7407 (2000).
\newline
30. G.R.Holt, J.Math.Phys. 23(6), 1037-1046 (1982).
\newline
31. G.Pucacco and Rosquist, J.Math.Phys. 46, 052902.1-21 (2005).
\newline
32. A.S.Fokas and P.A.Langerstrom, J.Math.Anal.Appl. 74, 325-341
(1980), 74, 342-358 (1980).
\newline
33. J.Hietarinta, Physics Reports 147, 87-154 (1987).
\newline
34. G.Thompson, J.Math.Phys. 25(12), 3474-3478 (1984).
\newline
35. J.Hietarinta, Phys.Lett. A246, 97-104 (1998), J.Math.Phys. 25,
1833-1840 (1984).
\newline
36. S.Gravel and P.Winternitz, J.Math.Phys. 43(12), 5902 (2002).
\newline
37. M.B. Sheftel, P.Tempesta and P.Winternitz, J.Math.Phys. 42(2),
659-673 (2001).
\newline
38. S.Gravel, J.Math.Phys. 45(3), 1003-1019 (2004).
\newline
39. P.A.Clarkson and M.D.Kruskal, J.Math.Phys. 38(11), 5944-5958
(2007).
\newline
40. D.Levi and P.Winternitz, J.Phys. A22, 2915-2924 (1989).
\newline
41. E.L.Ince, Ordinary Differential Equations (Dover, New York,
1944).
\newline
42. M.J.Ablowitz and P.A.Clarkson, Solitons, Nonlinear Evolution
Equations and Inverse Scattering, (Cambridge Univ.Press, 1991).
\newline
43. W.I.Fushchych and A.G.Nikitin, J.Math.Phys. 38(11) 5944-5959
(1997).
\newline
44. A.P.Veselov and A.Shabat,Funkt.Analiz.Prilozh 27(2), 1-21
(1993).
\newline
45. A.P.Veselov, J.Phys. A34, 3511 (2001).
\newline
46. I.Marquette and P.Winternitz, J.Math.Phys. 48(1) 012902
(2007).
\newline
47. D.Bonatsos, C.Daskaloyannis and K.Kokkotas, Phys.Rev. A48,
R3407-R3410 (1993).
\newline
48. C.Daskaloyannis and K.Ypsilantis, J.Math.Phys 47, 042904.1-38
(2006).

\end{flushleft}
\end{document}